\documentclass[prl,twocolumn,showpacs,preprintnumbers,amsmath,amssymb]{revtex4}

\usepackage{graphicx}
\usepackage{dcolumn}

\newcommand{\beq}{\begin{equation}}
\newcommand{\eeq}{\end{equation}}
\newcommand{\beqa}{\begin{eqnarray}}
\newcommand{\eeqa}{\end{eqnarray}}

\begin{document}
\title{Phase Separation close to the density-driven Mott transition in the
Hubbard-Holstein model}

\author{M. Capone$^{1,2}$, G. Sangiovanni$^2$, C. Castellani$^2$,
C. Di Castro$^2$, and M. Grilli$^2$} 

\affiliation{$^1$Enrico Fermi Center, Rome, Italy}

\affiliation{$^2$Istituto Nazionale di Fisica della Materia, 
Unit\`a Roma 1 and SMC Center, and Dipartimento di Fisica\\
Universit\`a di Roma "La Sapienza" piazzale Aldo Moro 5, I-00185 Roma, Italy}

\begin{abstract}
The density driven Mott transition is studied by means of Dynamical
Mean-Field Theory in the Hubbard-Holstein model, where the Hubbard term
leading to the
Mott transition is supplemented by an electron-phonon (e-ph) term. 
We show that 
an intermediate e-ph coupling leads to a first-order transition 
at $T=0$, which is accompanied by phase separation between a metal and an 
insulator. The compressibility in the metallic phase is substantially 
enhanced. At quite larger values of
the coupling a polaronic phase emerges coexisting with a non-polaronic metal.
\end{abstract}
\date{\today}
\maketitle

Since long time it is known that strong electron-electron (e-e) interactions
can drive a metallic system insulating \cite{mott}.
The correlation caused Metal-Insulator transition (MIT),
usually known as the Mott transition, occurs in several
compounds like $V_2O_3$, $CaTiO_3$
and is often theoretically investigated in the framework of the
Hubbard Hamiltonian. However, although this latter certainly captures
the fundamental properties of correlated electronic systems,
it lacks the lattice degrees of freedom, which can play a crucial
role in many respects. Therefore, not only the interplay between
the electronic correlation and the electron-phonon (e-ph) interaction is
a relevant issue in general \cite{cg,MILLIS},
but interesting specific effects can arise from it, like
in doped fullerenes, where a substantial 
e-ph coupling associated to Jahn-Teller distortion
can be synergetic with the large electron-electron
interaction to produce the observed rather large superconducting
temperatures \cite{capone}. Many experimental results also point
toward an important role of phonons in the strongly correlated cuprates.
In this light, it has also been proposed that the e-ph coupling 
could induce charge instabilities in the cuprates \cite{cg,cdg} thereby 
relating the interplay between strong e-e and e-ph interactions to the
 issue of electronic phase separation (PS), stripe formation, and to
 the isotopic dependence of the pseudogap crossover temperatures
 in these materials \cite{andergassen}.
 This connection between e-ph coupling and charge instabilities
 suggest that lattice degrees of freedom may greatly influence the
 thermodynamical stability of a strongly correlated
 electron system. 
 The present work precisely focuses on the role of
 the e-ph coupling in modifying or emphasizing the physical properties
 of electrons close to their Mott insulating status. Specifically we study
 the role of phonons in changing the order of the density-driven
MIT and in inducing PS close to the transition.

From the technical point of view
 we exploit the great progress made in this field with the dynamical
 mean-field theory (DMFT), which emerged as the 
first theoretical tool able to give a complete characterization of the
Mott transition\cite{revdmft}. This technique, which becomes exact 
in the infinite coordination limit\cite{metzner}, is in fact able
to treat with the same accuracy both the metallic and the insulating phase,
and it has been successfully applied to the Mott transition
in the  Hubbard model. Moreover, DMFT allows to treat on the 
same footing the e-e and e-ph interactions, which is crucial to 
our study. The main drawback of this approach is the 
momentum-independence of the self-energy, and the neglect of 
non-local interactions. This is not expected to 
introduce significant biases when the interactions are local, 
like in the model we study. 

One of the most striking features of the DMFT treatment of the Mott
transition is that both the interaction-driven transition at 
half-filling\cite{georges,kotliar}
and the density-driven transition\cite{fisher} are characterized by 
a coexistence of solution in some region of parameters. 
In the latter case, which we address in this paper, there
is a region of chemical potentials, delimited by two curves $\mu_{c1}(U)$ and
$\mu_{c2}(U)$,  in which a metallic solution
with $n\neq 1$ and an insulating one with $n=1$ coexist\cite{fisher}. 
Naturally, the stable solution is the one that minimizes the
grand-canonical free energy.
At $T=0$ the metal is stable in the whole coexistence region, and 
is continuously connected to the insulator \cite{fisher}. 
Moreover, the metal joins onto the insulator with a finite
slope in the $n$-$\mu$ curve, i.e., it has a finite 
compressibility $\kappa = \partial n/\partial\mu$ \cite{fisher,kajueter}.
At finite temperature the energetic balance gets more involved, and 
a first-order transition occurs along an intermediate line between 
$\mu_{c1}$ and $\mu_{c2}$, leading to PS between the 
two phases\cite{sahana}.

In this work we study how this scenario is modified by the inclusion
of a short-range interaction with phonons.
We consider the e-ph interaction in 
its simplest model realization, the Holstein molecular crystal model,
in which tight-binding electrons interact with local modes of constant 
frequency\cite{holstein}. The Hamiltonian is 
\begin{eqnarray}
H = & -t\sum_{\langle i,j\rangle}  c^{\dagger}_i c_j + H.c.
+ U \sum_i n_{i\uparrow} n_{i\downarrow} \nonumber\\
& - g\sum_i n_i(a_i +a^{\dagger}_i) + \omega_0 \sum_i a^{\dagger}_i a_i,
\label{hamiltonian}
\end{eqnarray}
where $c_i$ ($c^{\dagger}_i$) and $a_i$ ($a^{\dagger}_i$) are, respectively,
 destruction (creation) operators for 
fermions and for local vibrations of frequency $\omega_0$ 
on site $i$, $t$ is 
the hopping amplitude, $g$ is an e-ph coupling.

Previous studies have identified two different dimensionless e-ph
couplings, $\lambda=2g^2/\omega_0 t$, which measures
the energetic convenience to form a bound state, and $\alpha=g/\omega_0$
which controls the number of excited phonons. The relevance of each coupling
depends on the adiabatic ratio $\gamma = \omega_0/t$: If $\gamma$ is small
$\lambda$ is the control parameter, while for large $\gamma$ the physics is
mainly controlled by $\alpha$\cite{csg}.
Here we work with $\gamma =0.2$, and 
therefore we use $\lambda$ to measure the strength of the
e-ph coupling.

In DMFT, the lattice model is mapped onto an impurity
problem subject to a self-consistency condition, which contains
all the information about the lattice.
In the case of the Hubbard-Holstein model we have an Anderson impurity 
with a local phonon on the impurity site.
If we work in the Bethe lattice of half-bandwidth $t$
the self-consistency enforcing the DMFT solution is given by
\begin{equation}
\label{self}
\frac{t^2}{4}G(i\omega_n) = \sum_k \frac{V_k^2}{i\omega_n - \epsilon_k},
\end{equation}
where $\epsilon_k$ and $V_k$ are the energies and the 
hybridization parameters of the Anderson impurity model.
We use exact diagonalization to solve the impurity
model \cite{caffarel}. The technique  
consists in restricting the sum in Eq. (\ref{self}) to a finite and
small number of levels $N_s -1$. 
The discretized model is then solved using the
Lanczos method, which allows to compute the Green's function at $T=0$.  
The method converges rapidly as a function of $N_s$, and just 
a few levels are enough to achieve convergence. 
Here we mainly present results for $N_s = 10$, 
having checked in specific cases that no measurable change occurs in the
physical quantities by increasing $N_s$\cite{note-ph}.

As anticipated above, we discuss how the density-driven
Mott transition is affected by the e-ph interaction.
The main trend, which can be anticipated on intuitive grounds, is 
that the e-ph interaction favors the insulating phase
with respect to the metal. We note that in Ref. \cite{sahana}
the same effect is obtained by raising the temperature, because in 
that case the insulating state has a larger entropy. 
Here, when the e-ph interaction is turned on in the
metallic state, two competing effects take place. On one side, the
electron gains potential energy if the lattice is distorted, while,
on the other side, the effective mass is enhanced, leading to a
loss in kinetic energy.
As a result of this competition, the electrons can not 
completely exploit the e-ph coupling to lower the energy.
On the other hand, the electrons in the Mott insulator are already localized,
so that they can gain potential energy by coupling with phonons 
without losing kinetic energy.
This simple argument already tells us that the energetic balance between 
the two phases will be changed in favor of the insulator.
For the same reasons we can also expect the region of $\mu$ where the 
insulator exists to be larger, i.e., that $\mu_{c1}$ decreases with
increasing $\lambda$ \cite{notadoping}.

We notice that simplified
approaches, as the one based on Lang-Firsov and squeezing
transformations\cite{feinberg} also suggest that the insulating 
behavior may be favored by the electron-phonon interaction.
Within this approach, the Hubbard-Holstein model is transformed into 
a  Hubbard model with reduced hopping, in which the insulating behavior
is clearly favored with respect to the case where the electron-phonon 
interaction is absent.

\begin{figure}[htbp]
\begin{center}
\includegraphics[width=8.5cm]{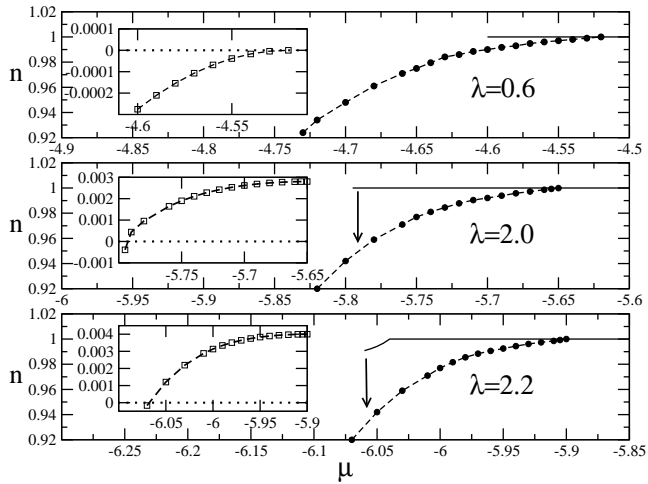}
\end{center}
\caption{The $n$-$\mu$ curves for $U/t=10$ for $\lambda = 0.6, 2, 2.2$. 
Solid circles joined by a dashed line denote the metallic solutions, 
and the solid line marks the insulating one. The insets show the difference between 
the grand canonical potentials of the two phases $\Delta\Omega = \Omega_{met}-
\Omega_{ins}$. In the two lower panels this quantity is negative signaling
the stability of the insulator. An arrow marks the transition point on the
$n$-$\mu$ curves}
\label{nvsmu}
\end{figure}

\begin{figure}[htbp]
\begin{center}
\includegraphics[width=7.5cm]{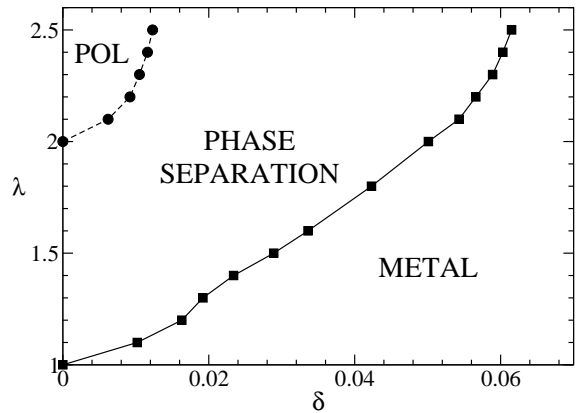}
\end{center}
\caption{The phase diagram for $U/t=10$ as a function of doping 
$\delta$ ($n=1-\delta$) and $\lambda$.} 
\label{phd}
\end{figure}

These naive arguments are confirmed and based on solid ground by
the DMFT calculations.
We have studied in detail both the solutions as a function of $\lambda$ 
for values of $U/t$ larger than the critical value 
$U/t \simeq 3$ for the Mott transition at half-filling \cite{revdmft}.
In Fig. \ref{nvsmu} we plot the density as a function of the chemical
potential for $U/t=10$ and various values of $\lambda$. For $\lambda = 0$ 
(not shown) we recover the known results for the Hubbard model, 
with two solutions in a really small $\mu$ interval, and the metal is
stable in the whole coexistence region, as it can be checked by comparing
the grand canonical potentials $\Omega = E -\mu N$ of the two phase. 
In the insets of the various panels we plot the difference $\Omega_{met} -
\Omega_{ins}$.
For small values of $\lambda$ we observe a slight
increase of the coexistence region, but the metal stays at lower energy than 
the insulator (see upper panel of Fig. \ref{nvsmu}). 
For $\lambda > 1$ the two curves start to cross and the grand canonical
potential of the insulator becomes lower than the one of the metal
for some range of chemical potential 
(in the middle panel we show the result for $\lambda = 2$), 
leading to a jump of the occupation number by continuously varying the 
chemical potential.
The transition becomes therefore of first order, and the chemical potential
has a plateaux with respect to the density. As a result, the system is
not stable in the interval of density between which  the jump occurs, and it
undergoes PS between the insulating solution with $n=1$ 
and a metallic solution with some density $n_{ps}(\lambda)$.
$n_{ps}$ is found to be a decreasing function of $\lambda$.
In Fig. \ref{phd} we show the PS region for $U/t= 10$.

If we further increase the coupling, the effect of the e-ph
interaction becomes even more dramatic. In the case of $U/t=10$, and 
$\lambda \simeq 2.2$, the insulating solution continuously evolves into 
a phase with density slightly smaller than 1. Also for this coupling
the insulator is energetically favored in the coexistence region
(see instet of lower panel in Fig. \ref{nvsmu}). 
An inspection to other physical quantities allows us to better characterize
this strong-coupling phase as a polaronic state. 
In fact, if we compute the phonon displacement distribution function
$P(X) = \langle X\vert\psi_0 \rangle  \langle\psi_0 \vert X\rangle$,
the evolution from the $n=1$ insulator to this state is accompanied
by a change from a unimodal distribution of the phonon displacement to
a bimodal distribution characteristic of the polaronic state.
On the other hand, the metallic solution does not display polaronic features 
for these values of $\lambda$ and so the PS in this region,
takes place between a polaronic insulator and a non-polaronic metal.
 
Our DMFT results show that the Hubbard-Holstein model displays 
PS close to half-filling, due to the first-order phase transition
between the insulator and the metal.
Even if PS occurs close to half-filling, just like in the large$-N$ 
treatment of the same model, the present
result is physically quite different. In Ref. \cite{becca}
a divergent compressibility was indeed found within the metallic phase.
Here this instability is prevented by the stabilization of the insulator.
Nevertheless, we can follow the metallic solution in the regime where
it is metastable, and check whether the e-ph interaction
may lead to a divergence, or at least to a sizeable enhancement of the
charge compressibility.

In the infinite-$U$ limit, the large$-N$ approach has shown that the charge 
compressibility is well represented by a simple RPA-like formula,
\begin{equation}
\label{rpa1}
\kappa (\lambda)= \frac{2N^*}{1+(F_0^S)_{e} + (F_0^S)_{ph}}, 
\end{equation}
where $\kappa(\lambda)$ is the compressibility for a given value of 
the e-ph coupling $\lambda$, $N^*$ is the quasiparticle
density of states per spin at the Fermi level, $(F_0^S)_{e(ph)}$
is the electronic
(phononic) contribution to the symmetric Landau amplitude, and $(F_0^S)_{ph}
= -4N^*g^2/\omega_0$, which differs
from our $\lambda$ for the presence of the quasiparticle DOS instead of
the real DOS. From (\ref{rpa1}) it is easy to derive
$\kappa (\lambda)/ \kappa(0) = 1/(1-\lambda\kappa(0)t)$, 
in which all the correlation effects are contained in the 
purely electronic compressibility calculated at $\lambda=0$.
This implies that, if the electronic compressibility
is large, a small or moderate e-ph coupling is sufficient to make
the system unstable in the charge channel.

\begin{figure}[htbp]
\begin{center}
\includegraphics[width=7cm,height=6cm]{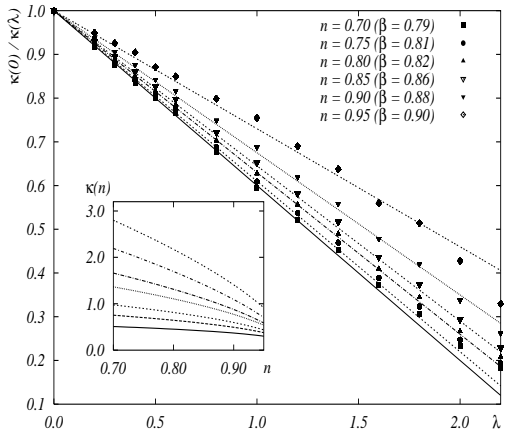}
\end{center}
\caption{Normalized inverse compressibility
$\left[ \kappa(\lambda)/\kappa(0)\right]^{-1}$
as a function of $\lambda$ at $U/t=5$ and 
for different densities, compared with the fit
 described in the text. Inset: Compressibility
vs filling for various values of $\lambda=0.0, 0.8, 1.2, 1.6, 1.8, 2.0, 2.2$
(from the bottom to the top).
}
\label{fit}
\end{figure}

Coming back to the results shown in Fig. 1, the 
$n$-$\mu$ curves of the metallic solutions have a small finite slope when 
they approach $n=1$. However, if we consider the compressibility 
at fixed density as a function of $\lambda$, it turns out that
the e-ph interaction  is substantially increasing
the compressibility.  
In Fig. \ref{fit} we present the inverse of the normalized compressibility
as a function of $\lambda$ at $U/t = 5$ for different values of the electron
density, ranging from $n=0.70$ to $n=0.95$. An enhancement of
$\kappa(\lambda)/\kappa(0)$ that varies from almost 6 in the $n=0.70$
case, to around 3 in the $n=0.95$ case corresponds to the decreasing linear
behavior of all the curves seen in the figure.
We also calculated the values of the renormalization factor $Z$ which is
inversely proportional to the quasiparticle density of states.
Since we found that $Z$ varies much less than
$\kappa$ as a function of $\lambda$, such an enhancement of the
compressibility cannot be due to mass renormalization alone.
This also indicates that, within the Landau
Fermi liquid approach, the total $F_0^S$ is expected to be renormalized 
by the e-ph interaction like in the large$-N$
calculations.
We extract informations on this renormalization of the Landau
amplitudes by simply assuming that a relation similar to 
Eq. (\ref{rpa1}) holds also for the DMFT solution.
We therefore fitted the DMFT data with the relation
\begin{equation}
\frac{\kappa(\lambda)}{\kappa(0)} = 
\frac{1}{1-\beta\lambda\kappa(0)t},
\label{ourfit}
\end{equation}
where $\beta$ is the only fitting parameter. 
The results of this fit, together with the values of $\kappa(\lambda)$
in the inset, 
are shown in Fig. \ref{fit}. 
For almost the whole range of densities the fit is extremely accurate up to
$\lambda = 2.2$, with just a small correction to the ideal large-N
result $\beta =1$.
The closer we get to half-filling the more $\beta$ approaches
1, even if at $n=0.95$ the strong-coupling data display some fluctuations.
The values of $\beta$ are reported in the figure.

In light of the ability of the simple Eq. (\ref{ourfit}) to describe
the way the e-ph interaction modifies the scenario
determined by electron correlation, the small values of 
$\lambda$ needed to obtain a divergent compressibility 
in the large$-N$ approach \cite{cg,becca}
 can be related to the smaller value of the
electronic $F_0^S$ found in the large$-N$ with respect to the one found in
the present DMFT approach.

Our findings could be related to some 
experimental observations. First of all the presence of
a (phonon-induced) PS close to a Mott
insulating phase is suggestive of a chemical potential plateaux
inferred from photoemission experiments in underdoped cuprates
\cite{ino}. The formation of doping-induced
metallic states with a chemical potential slowly varying with doping
inside a Mott gap was also recently deduced from tunneling
experiments \cite{bozovic}. In the same experiment, the small
mixing of the metallic (superconducting) phase with the antiferromagnetic
insulating one indicates that a spatial separation of the two phases
can have small interfaces suggesting that the presence of long-range 
Coulomb forces can easily induce small-scale (presumably stripe-like)
domain formation. 
We also remarkably found that for rather large values of the
e-ph coupling ($\lambda >2$ in the $U=10t$ case) a 
PS can occur between an insulator away from half-filling
with a finite density of polarons and a good metal with nearly free carriers.
Also this situation of coexisting (but spatially separated) polarons 
and free carriers could find a realization in underdoped cuprates
\cite{MUELLER}.

In conclusion, we have studied the effect of the e-ph 
interaction on the density-driven Mott transition for large values 
of $U$. The e-ph interaction favors the insulating 
solution with respect to the metal. This effect, besides a quantitative
modification of the coexistence region between the two solutions,
determines a first-order transition between an insulator and a metal at
some value of the chemical potential $\mu_c$, and to a 
PS between the Mott insulator with $n=1$
and a metal with a finite density $n \neq 1$.
This PS occurs for $\lambda \simeq 1$. For substantially
larger values of the e-ph coupling a polaron crossover occurs
very close to half-filling and PS takes place between
this polaronic state and a non-polaronic metal.

We acknowledge fruitful discussions with S. Ciuchi and 
support from MIUR Cofin 2001, prot. 2001023848, and from INFM, 
also through PA\_G0\_4 project.


\begin{thebibliography}{99}
\bibitem{note-ph} The phonon Hilbert space has also been truncated by 
allowing  for a finite number of phonons and checking for 
convergence as a function of this number. 
When necessary, we considered up to 50 phonons.
\bibitem{notadoping} We consider here hole doping $(n \le 1)$. Owing to the
particle-hole symmetry of the model, the same physical behavior is obtained for
electron-doping $(n\ge 1)$.
\end{thebibliography}
\end{document}